\documentclass[aps,prx,twocolumn,groupedaddress,floatfix,longbibliography]{revtex4-2}

\usepackage{graphicx}
\usepackage{dcolumn}
\usepackage{bm}
\usepackage{amsmath}
\usepackage{xcolor}
\usepackage{braket}
\usepackage[utf8]{inputenc}
\usepackage[T1]{fontenc}
\usepackage{hyperref}
\hypersetup{
breaklinks=true,
    colorlinks=true,
    linkcolor=blue,
    citecolor=blue,
    filecolor=magenta,
    urlcolor=cyan,
}

\usepackage{blkarray}
\usepackage{multirow}

\usepackage{natbib}      

\newcommand{\cc}{{\cal C}}

\newcommand{\bG}{{\bf G}}
\newcommand{\br}{{\bf r}}

\begin{document}

\title{Quantum Wigner molecules in moir\'{e} materials}

\author{Constantine Yannouleas}
\email{Constantine.Yannouleas@physics.gatech.edu}
\author{Uzi Landman}
\email{Uzi.Landman@physics.gatech.edu}

\affiliation{School of Physics, Georgia Institute of Technology,
             Atlanta, Georgia 30332-0430}

\date{Submitted: 19 April 2023; Letter, PRB {\bf 108}, L121411 (2023)}

\begin{abstract}
The few-body problem (with $N \leq 6$ fermionic charge carriers) in isolated moir\'{e} quantum dots 
(MQDs) in transition metal dichalcogenide (TMD) bilayer materials with integer fillings, $\nu \geq 2$, 
is investigated by employing large-scale full configuration interaction (FCI, also termed 
exact-diagonalization) computations, and by performing a comparative analysis of the ensuing 
first-order (charge densities, CDs) and second-order (conditional probability distributions, CPDs) 
correlation functions. 
\textcolor{black}{With parameters representative of bilayer experimental TMD setups,}
our investigations reveal the determining role of the strong inter-particle
Coulombic repulsion in bringing about Wigner molecularization, which is associated with many-body 
physics beyond both that described by the aufbau principle of natural atoms, as well as by the 
widely used Hubbard model for strongly-interacting condensed-matter systems. In particular,
for weak and moderate 
\textcolor{black}{trilobal crystal-field deformations} of the MQDs, the imperative employment 
of the CPDs brings to light the geometrical polygonal-ring configurations underlying the
Wigner molecules (WMs) that remain hidden at the level of a charge-density analysis, apart from the
case of $N=3$ when a {\it pinned\/} WM emerges in the charge density due to the coincidence of the
$C_3$ symmetries associated with both the intrinsic geometry of the $N=3$ WM 
\textcolor{black}{and the TMD trilobal crystal-field of the confining pocket potential.} 
\textcolor{black}{The FCI numerically exact-diagonalization results 
provide critical benchmarks for assessing and guiding the development of future computational 
methodologies of interacting strongly-correlated fermions in isolated MQDs and their superlattices 
in TMD materials.}
\end{abstract}

\maketitle

Understanding of the electronic spectral and configurational organization in natural atoms, which
played a pivotal role in the early development of quantum mechanics \cite{jamm,pais}, continues to
inspire discoveries in research targeting the exploration of the nature of few charged carriers trapped 
in artificially fabricated, isolated or superlattice-assembled, quantum dots (QDs) 
\cite{kouw97,hans07,jing22}. 
Such research aims at utilizing these systems, with high tunability and control, in future quantum 
information and computational platforms \cite{copp13,vand19,deng20,burk21}. Earlier studies have unveiled 
formation of quantum Wigner molecules (WMs), originally predicted theoretically 
\cite{yann99,grab99,yann00,fili01,yann02.2,mikh02,harj02,yann03,szaf03,yann04,roma06,yann06.3,yann07,yang07,
yann07.3,umri07,yang08,roma09,yann15,yann21,erca21.2,urie21,yann22,yann22.2}
in two-dimensional (2D) semiconductor QDs, as well as in trapped ultracold atoms,
and subsequently observed experimentally in GaAs QDs \cite{yann06,kall08,kim21,kim22}, Si/SiGe 
QDs \cite{corr21}, and carbon-nanotubes \cite{peck13}. Important contributions to this area
employed full configuration-interaction (FCI) 
\cite{shav98,yann03,szaf03,yann06.2,ront06,yann07,yann07.3,yann15,yann21,yann22.2,yann22.2,yann22.3,szabo}
calculations, going beyond (1) the central-field Aufbau principle, which underlies the 
periodic table of natural atoms \cite{ostr01}, 
and (2) the Hubbard modeling of strongly interacting systems \cite{hubb63,hubb78}. 

\textcolor{black}{
Here we broaden the above WM-portfolio, by uncovering, with the use of FCI calculations, the ubiquitous 
formation of WMs in a novel class of two-dimensional (2D) materials which emerged recently as a most 
promising
platform for investigations of strongly correlated electrons. These materials form moir\'{e} superlattices 
(with large, tunable lattice constants) when 2D crystals (made of semiconductor or semimetal materials) are 
stacked with a small twist angle or lattice mismatch; for references to the developing literature on twisted 
$\Gamma$-valley homo-bilayer (e.g., MoS$_2$), or hetero-bilayer (e.g., WSe$_2$/WS$_2$, MoSe$_2$/WSe$_2$) 
transition metal dichalcogenides (TMDs), and the emergent twistronics field, see 
\cite{manz17,kaxi20,macd18,fu20,ange21}.} 

\begin{figure*}[t]
\centering\includegraphics[width=15.0cm]{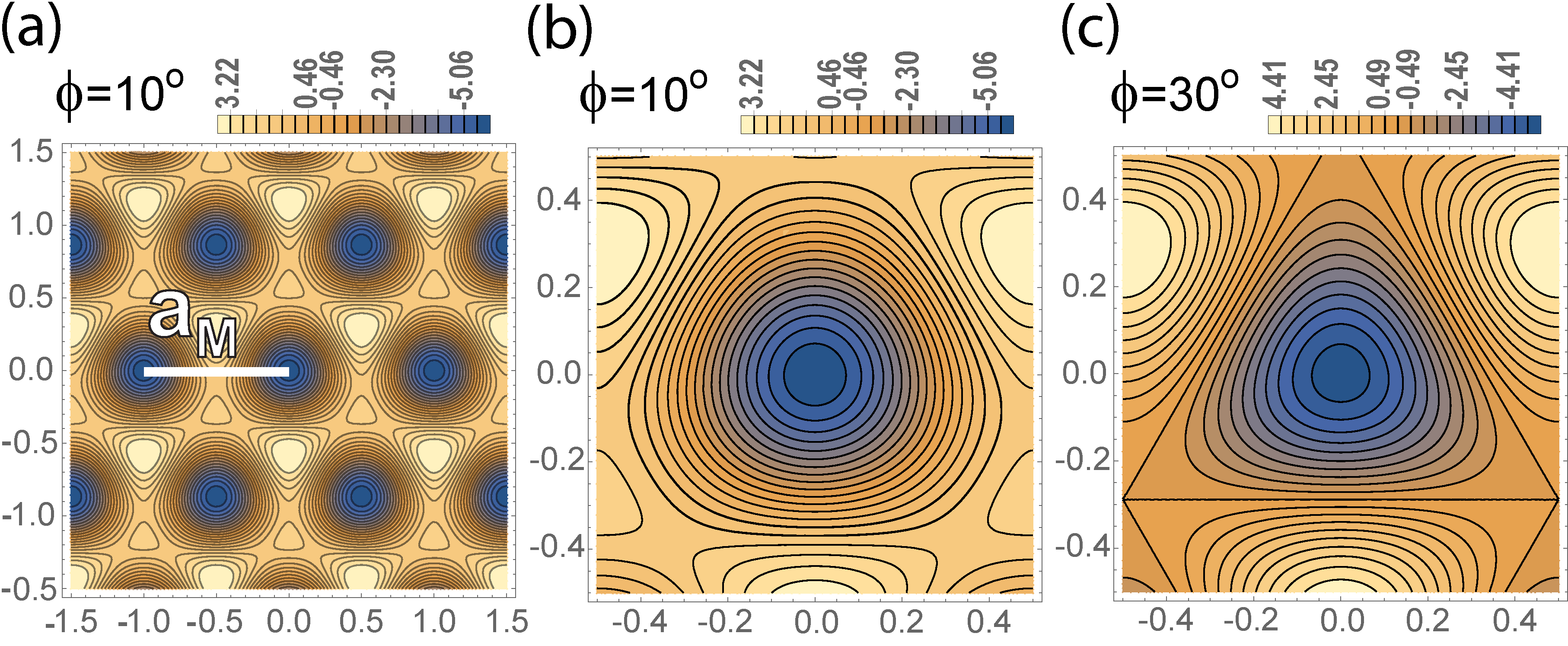}
\caption{
Plot of the moir\'{e}-superlattice potential given by Eq.\ (\ref{mpot}). (a) Broader view of the
periodic potential structure for an angle of $\phi=10^\circ$. (b) Potential of the isolated moir\'{e} QD for
$\phi=10^\circ$. (c)  Potential of the isolated moir\'{e} QD for $\phi=30^\circ$. Lengths in units of
the moir\'{e} lattice constant $a_M$. Potential contours in units of $v_0$.
Note the change of the length scale in (b) and (c) compared to (a).}
\label{potm}
\end{figure*}

We focus here on few-fermion ($N<7$, electrons or holes) moir\'{e} quantum dots (MQDs) 
\cite{feen18,zeng22,song22} formed at the upper layer of doped integer-filled bilayer TMDs 
\cite{mak21,feld22}. The potential confining these fermions at the minima of the 2D moir\'{e} superlattice 
\textcolor{black}{can be approximated by the expression \cite{macd18,ange21,fu20}}
\begin{align}
V(\br) = -2 v_0 \sum_{i=1}^3 \cos(\bG_i \cdot \br + \phi),
\label{mpot}
\end{align}      
where $\bG_i=[ (4\pi/\sqrt{3}a_M) ( \sin(2\pi i/3), \cos(2\pi i/3) ) ]$ are the moir\'{e} reciprocal 
lattice vectors. The materials specific parameters of $V(\br)$ are $v_0$ (which can also be experimentally 
controlled through voltage biasing), the moir\'{e} lattice constant $a_M$, and the angle $\phi$; for variations
of $V(\br)$ with $\phi$, see Fig.\ \ref{potm}; $a_M$ is typically of the order of 10 nm, which is much larger 
than the lattice constant of the monolayer TMD material (typically a few \AA). The parameter $\phi$ 
controls the strength of the trilobal crystal-field-type anisotropy in each MQD potential pocket.

\textcolor{black}{ 
For angles $\phi < 40^\circ$ and a large ratio $v_0/E_K >>1$ (where the moir\'{e} kinetic energy 
$E_K=\hbar^2/(2m^*a_M^2)$ \cite{fu20}, with $m^*$ being the effective mass of the charge carriers,
(referred to here, indiscriminately, as 'holes') the attractive pockets 
of $V(\br)$ represent a periodic array of isolated MQDs \cite{feen18,zeng22,song22} 
where the carriers are tightly bound.}
In a superlattice with integer fillings, the corresponding strongly-interacting few-body problem (with 
$N \leq 6$ holes) has already been experimentally realized \cite{mak21,feld22}).

\textcolor{black}{
A prerequisite to materialization of gate-controlled tunable MQDs in bilayer TMDs (in isolation or in 
superlattices) with integer fillings (via doping) in quantum information and simulations, is a 
thorough understanding of the correlated electronic states of the strongly-interacting MQD-confined  
charge carriers.}
To this aim we concentrate on TMD materials with a large lattice constant, $a_M$, where the 
QMDs are evocative of the electrostatically defined ones in 2D semiconductors (SQDs, e.g., in GaAs 
\cite{yann07}) with a major difference which is readily grasped by 
expanding $V(\br)$ in Eq.\ (\ref{mpot}) in powers of $r$, and defining a confining potential,
$V_{\rm MQD}(\br)$, for the isolated MQD as follows:
\begin{align}
\begin{split}
V_{\rm MQD}(\br) \equiv V(\br) + 6 v_0 \cos(\phi)  \approx m^* \omega_0^2 r^2/2 + \cc \sin(3 \theta) r^3.
\end{split}
\label{vexp} 
\end{align}
with $m^*=0.50 m_e$, $m^*\omega_0^2=16 \pi^2 v_0 \cos(\phi)/a_M^2$, and 
$\cc=16 \pi^3 v_0 \sin(\phi)/( 3\sqrt{3}a_M^3)$ (the expansion of $V(\br)$ can be restricted to the terms 
up to $r^3$). 
\textcolor{black}{
Because of the anisotropic second term, the $V_{\rm MQD}(\br)$ confinement [Eq.\ (\ref{vexp})] 
has a crystal-field, $C_3$ (trilobal) point-group symmetry and differs from that 
of the extensively studied single circular 
2D QD, which is well approximated solely by the harmonic term [first term in $V_{\rm MQD}(\br)$ above]. 
Furthermore, consideration of anisotropic confinements in SQDs have treated elliptic-in-shape distortions 
[using $\omega_{0x} \neq \omega_{0y}$ in place of a single $\omega_0$ and ${\cal C}=0$ in Eq.\ (\ref{vexp})], 
under the influence of applied magnetic fields, $B$ \cite{yann07.2,urie21,yann07,szaf04}, lacking a 
comprehensive study of the trilobal anisotropy in QDs for $B=0$ (case of the MQD where diamagnetic effects 
are negligible due to large values of $\hbar \omega_0$).
}

Below, we analyze (using representative materials' parameters corresponding to experimentally investigated
and theoretically modeled TMD materials \cite{macd18,fu20,zeng22,feld22})
the emergence of quantum moir\'e WMs (MWMs) in the above trilobal confinement.
Specifically, our FCI results will show that a relatively weak 
$\sin(3 \theta) r^3$ term (case of $\phi < 15^\circ$) can act as a pinning agent for the WM only in the case 
of $N=3$, when the intrinsic WM azimuthal geometry [an empty-center (0,3) polygonal ring] coincides
with the $C_3$ point-group symmetry associated with $\sin(3\theta)$; in this pinned case, the WM is 
visible in the charge density. In all other instances investigated here, however, there is no pinning
\cite{[{Our findings concerning the prerequisites for WM pinning agree with the results in }] [{}]szaf09}, 
i.e., whilst the charge densities (CDs) preserve the $C_3$ symmetry (as quantum-mechanicaly required), they 
do not exhibit any signature of Wigner molecularization. Unveiling (seeing) the hidden (unseen) Wigner 
molecularization necessitates analysis which reaches beyond the information given by the CD distribution 
in the MQD; that is, it requires analysis of spin-resolved  density-density, or conditional probability 
distributions, (CPDs, see, e.g.,  Refs.\ \cite{yann00,yann04,yann06.3,yann07,yann15,yann09};
for the definition, see Appendix \ref{a2}), which reveal the intrinsic $(n_1,n_2)$ 
(with $n_1+n_2=N$, $N=2-6$) concentric polygonal-ring configurations, familiar from the literature
\cite{yann99,yann00,yann02.2,mikh02,yann03,szaf03,yann04,yann06.3,yann07,
yang07,umri07,yang08,yann22,yann22.2} on circular SQDs.

\begin{figure}[t]
\centering\includegraphics[width=7.8cm]{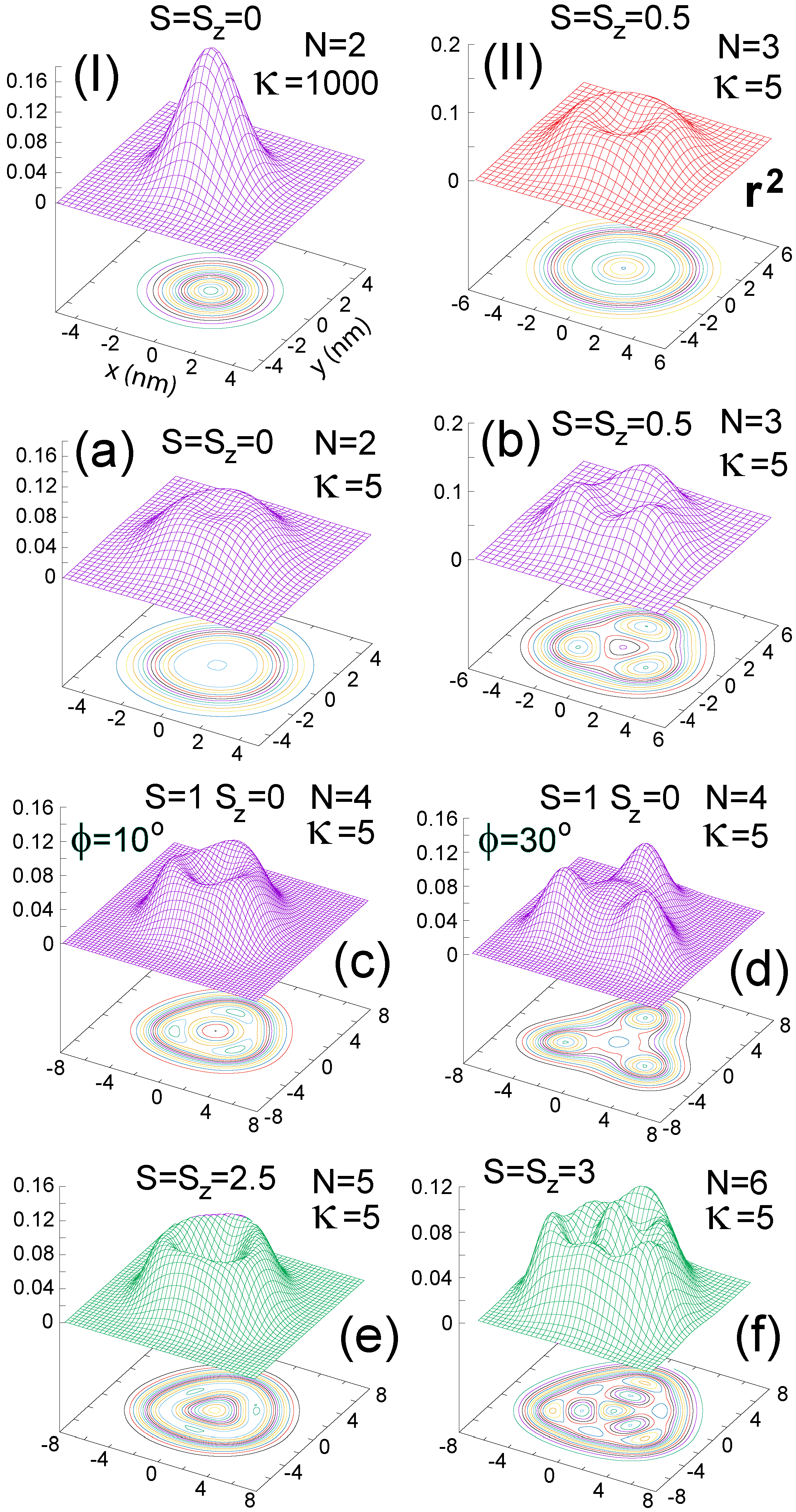}
\caption{
FCI CDs (normalized to $N$) for $N=2-6$ holes in the isolated MQD. In all frames: $a_M=14$ nm, 
$m^*=0.5m_e$, $\phi=10^\circ$, $v_0=15$ meV, resulting in $\hbar \omega_0=42.59073$ meV 
\textcolor{black}{[but $\phi=30^\circ$ in (d)].}
$\kappa=5.0$ ($R_W=3.57$) except in frame (I), where $\kappa=1000$ ($R_W=0.018$) mimicks the 
non-interacting limit. In all cases, the total anisotropic confinement was considered, except in frame (II) 
(red-colored surface) where the confinement consists solely of the 2D circular harmonic contribution. The 
green color denotes fully polarized ($S=S_z=N/2$) cases. A vanishing magnetic field, $B=0$, was considered, 
except for $N=3$ and $N=5$ where a value of $B=1$ T was used in order to simply lift the ground-state 
degeneracy. Frames for non-fully-polarized cases describe low-spin ground states. CDs in units of $1/nm^2$. 
See the text for a full description.}
\label{dens}
\end{figure}

The Schr\"{o}dinger equation for the many-body Hamiltonian of the isolated moir\'{e} QD, given by
\begin{align}
H_{\rm MB} = \sum_{i=1}^N \left\{ \frac{{\bf p}_i^2}{2 m^*} +V_{\rm MQD}(\br_i) \right\} +
\sum_{i<j}^N \frac{e^2}{\kappa |\br_i-\br_j|},   
\label{mbh}
\end{align}
is solved here using the FCI methodology 
\cite{shav98,yann03,szaf03,ront06,yann07,yann22.2,yann22.3,szabo};
for a brief description of this methodology, see Appendix \ref{a1}, including Ref.\ \cite{kime93}.
The emergence of MWMs is analyzed using both the FCI CDs and CPDs. $\kappa$ is the
\textcolor{black}{
effective dielectric constant obtained as geometric mean of the anisotropic in-plane and perpendicular 
tensor components of the MQD TMD dielectric environment (most often \cite{macd18,zeng22,feld22}
hexagonal boron-nitride, hBN, when $\kappa \approx 5$). 
We remark that because the spin and valley degrees of freedom are locked for the holes in TMD materials 
\cite{fu20,xiao12,macd18,macd23} only the spin needs to be considered in the course of the FCI 
exact-diagonalization of $H_{\rm MB}$. 
}

\begin{figure}[t]
\centering\includegraphics[width=7.7cm]{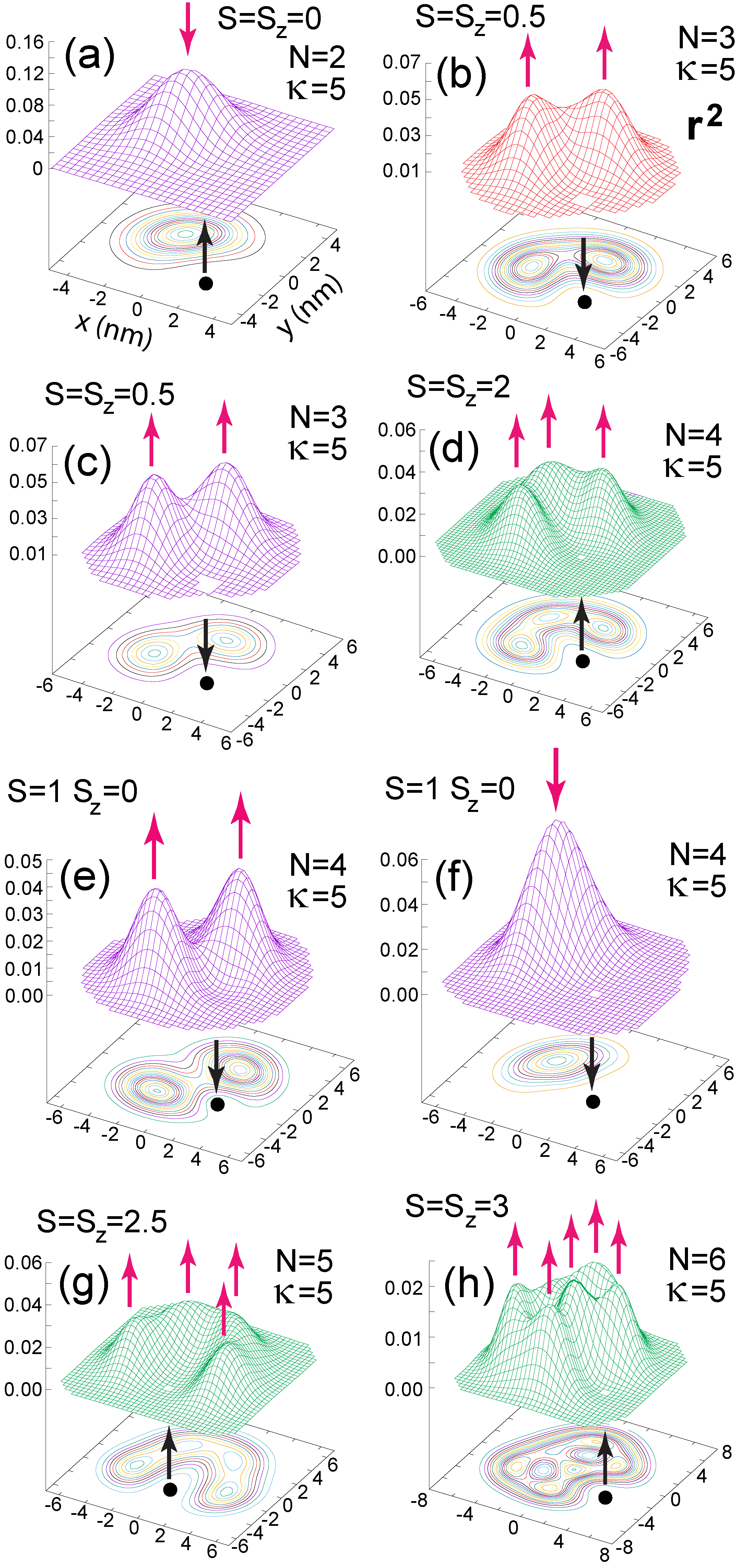}
\caption{
FCI spin-resolved CPDs (normalized to unity) for $N=2-6$ holes in the isolated MQD. Parameters as denoted
(compare Fig.\ \ref{dens}). $a_M=14$ nm, $m^*=0.5m_e$, $\phi=10^\circ$, $v_0=15$ meV.
In all cases, the total anisotropic $V_{\rm MQD}(\br)$ was considered, except in (b) 
(red-colored surface) where the confinement consists solely of the 2D circular harmonic contribution. The 
green color denotes again fully polarized ($S=S_z=N/2$) cases, and all the remaining frames describe low-spin 
ground states. The black solid dot and black arrows denote the position and spin direction of the fixed-point 
fermions, respectively. The red arrows denote the positions (maxima of the spin surfaces) and spin direction 
of the remaining $N-1$ fermions. $B=0$, except for $N=3$ and $N=5$ where a value of $B=1$ T was used in order 
to simply lift the ground-state degeneracy. CPDs in units of $1/nm^2$. See the text for a full description.}
\label{cpd}
\end{figure}

For values of $\phi=10^\circ$ and $\phi=30^\circ$, $V_{\rm MQD}(\br)$ exhibits, respectively, weak and 
strong $\sin(3\theta)r^3$ distortions away from the circular confinement ($\phi=0$); see illustration in 
Fig.\ \ref{potm}. In Figs.\ \ref{dens} and \ref{cpd}, we analyze primarily the 
$\phi=10^\circ$ case, \textcolor{black}{with an exception for $\phi=30^\circ$ in Fig.\ \ref{dens}(d).} 
Additional results for $\phi=30^\circ$, as well as $\phi=0$ and $\phi=10^\circ$, are presented 
in Appendix \ref{a3} and Appendix \ref{a4}.
The Wigner-molecularization propensity uncovered from the FCI results is controlled by the 
Wigner parameter $R_W =e^2/(\kappa l_0 \hbar \omega_0)$ \cite{yann99,yann07}, expressing the ratio between
the Coulomb repulsion and the quantum kinetic energy (proportional to the harmonic-oscillator 
energy gap); $l_0=[\hbar/(m^* \omega_0)]^{1/2}$ is the oscillator length. A high Wigner molecularization 
propensity is expected for $R_W > 1$.

Fig.\ \ref{dens} displays CDs for $N \leq 6$ in various instances for holes when $\phi=10^\circ$.
For contrast, Fig.\ \ref{dens}(d) employs $\phi=30^\circ$ and in Fig.\ \ref{dens}(II) only the circular 
partial $r^2$ term has been employed in $V_{\rm MQD}(\br)$. 
In Fig.\ \ref{dens}(I), $\kappa=1000$ was used to simulate the non-interacting limit, whereas $\kappa=5$ 
(strongly-interacting regime appropriate for the MQD in an hBN environment) was used in the rest of the 
frames.  

The two frames for $N=2$ [with $S=S_z=0$, Fig.\ \ref{dens}(I) and Fig.\ \ref{dens}(a)] demonstrate the 
effect of an increasing interparticle Coulomb repulsion (decreasing $\kappa$) on the CDs. 
Indeed, in Fig.\ \ref{dens}(I) (with $\kappa=1000$, non-interacting regime) the CD is  
associated with a doubly occupied $1s$ orbital. In 
contrast, for $\kappa=5$ (MQD/hNB case), the $N=2$ CD in Fig.\ \ref{dens}(a) is spread 
out over a considerably larger area and it assumes a ring-like shape by developing a depression at the 
origin. This ring-like shape preserves the $C_3$ symmetry of the $V_{\rm MQD}(\br)$ by being slightly 
deformed. 

The CD in Fig.\ \ref{dens}(a) does not exhibit the anticipated azimuthal geometry of a 
diatomic molecule, namely two localized particles in an antipodal arrangement. This WM configuration 
remains hidden in the CD of Fig.\ \ref{dens}(a), but it is revealed in the spin-resolved
CPD displayed in Fig.\ \ref{cpd}(a). Indeed, for a spin-up fixed hole, the second hole is found in an 
antipodal position with an opposite spin, as is consistent for a singlet state $(S=S_z=0)$.  

For $N=3$, the CDs in Fig.\ \ref{dens}(II) and Fig.\ \ref{dens}(b) exhibit different behaviors.
Specifically, when only the isotropic $r^2$ term is
retained in Eq.\ (\ref{vexp}), the CD is an azimuthally uniform ring [Fig.\ \ref{dens}(II)].
However, further inclusion of the anisotropic $\sin(3\theta)r^3$ term in $V_{\rm MQD}(\br)$ results 
in a drastic change; the CD becomes that of three holes localized at the apices of 
an equilateral triangle [Fig.\ \ref{dens}(b)]. One then can talk of a profoundly quantum, ``rotating'' 
(0,3) intrinsic WM polygonal-ring structure \footnote{Indeed, the term ``rotating Wigner (or electron) 
molecule'' has been used to accentuate the quantum nature of the Wigner molecule in earlier literature; 
see, e.g., Refs.\ \cite{yann03,yann04,yann06.3,yann07,yang07,yang08}}
in the case of Fig.\ \ref{dens}(II), which transforms into a {\it pinned\/} WM in Fig.\ \ref{dens}(b);
the intrinsic (0,3) molecular structure is {\it unseen\/} in the CD of the 
isotropic confinement [Fig.\ \ref{dens}(II)], but it becomes visible in the spin-resolved CPD in Fig.\ 
\ref{cpd}(b). For completeness, we also display [Fig.\ \ref{cpd}(c)] the spin-resolved CPD for $N=3$ 
associated with the total potential, which is redundant, however, since, in this case, the (0,3) 
molecular configuration is already revealed in the CD [Fig.\ \ref{dens}(b)].

With same parameters, the CDs for $N=4$ (ground state), and for $N=5$ and $N=6$ (lowest-in-energy fully 
spin-polarized states) are displayed in Fig.\ \ref{dens}(c), Fig.\ \ref{dens}(e), and Fig.\ \ref{dens}(f),
respectively. These three densities preserve the $C_3$ symmetry of the external confinement, but they 
are similar in shape and do not reveal any intrinsic molecular structure. Again, the intrinsic $(n_1,n_2)$ 
polygonal configurations are revealed through the FCI CPDs. In particular, two spin-resolved CPDs [spin-down 
at fixed-point, look for the two spin-ups, Fig.\ \ref{cpd}(e) and spin-down at fixed-point, look for the other 
spin-down, Fig.\ \ref{cpd}(f)], associated with the $N=4$ ground-state density in Fig.\ \ref{dens}(c), are 
presented. Taking into account the two remaining CPDs (spin-up, look for spin-up and spin-up, look for 
spin-down, not shown), one concludes that, in this particular case, the intrinsic spin eigenfunction of the 
rotating WM has the form 
$(|\uparrow\downarrow\uparrow\downarrow\rangle - |\downarrow\uparrow\downarrow\uparrow\rangle)/\sqrt{2}$
\footnote{For the complete set of possible $N=4$ spin eigenfunctions, see Ref.\ \cite{yann09}, 
and references therein},
with the localized fermions arranged into a distorted (0,4) polygonal configuration. 

In the case of the fully polarized states in Fig.\ \ref{cpd}(d) ($N=4$), Fig.\ \ref{cpd}(g) ($N=5$), and Fig.\
\ref{cpd}(g) ($N=6$), there is only one spin-resolved CPD (spin-up, look for spin-up); in all three cases, 
this CPD reveals directly the intrinsic structure of a (0,4), (0,5), and (1,5) polygonal WM, respectively.
These intrinsic polygonal configurations are necessarily slightly distorted in order to guarantee the $C_3$ 
symmetry of the charge densities; the CDs equal the integral of the CPDs over all possible positions of the 
fixed point  
\cite{ [{The subtle interplay discussed here between the {\it symmetry-preserving\/} exact CDs and the
apparently {\it symmetry-breaking\/} intrinsic WM configurations has been noted and discussed in earlier
literature; see Ref.\ \cite{yann07}, and references therein, Ref.\ \cite{yann02.2}, and }]
[{, and references therein}]shei21}.

\textcolor{black}{
Finally, the CD in Fig.\ \ref{dens}(d) for $\phi=30^\circ$ illustrates that increasing the strength of the  
crystal-field-like trilobal anisotropy induces a structural isomeric transition \footnote{For a comprehensive
study of the dependence of the WM on $\phi$ (the strength of the trilobal anisotropy), see C. Yannouleas 
and U. Landman, to be published.}
of the $N=4$ MQD from a ``rotating'' (or gliding) (0,4) WM [Fig.\ \ref{dens}(c)] to a pinned (1,3) one.
}

\begin{figure}[t]
\centering\includegraphics[width=7.8cm]{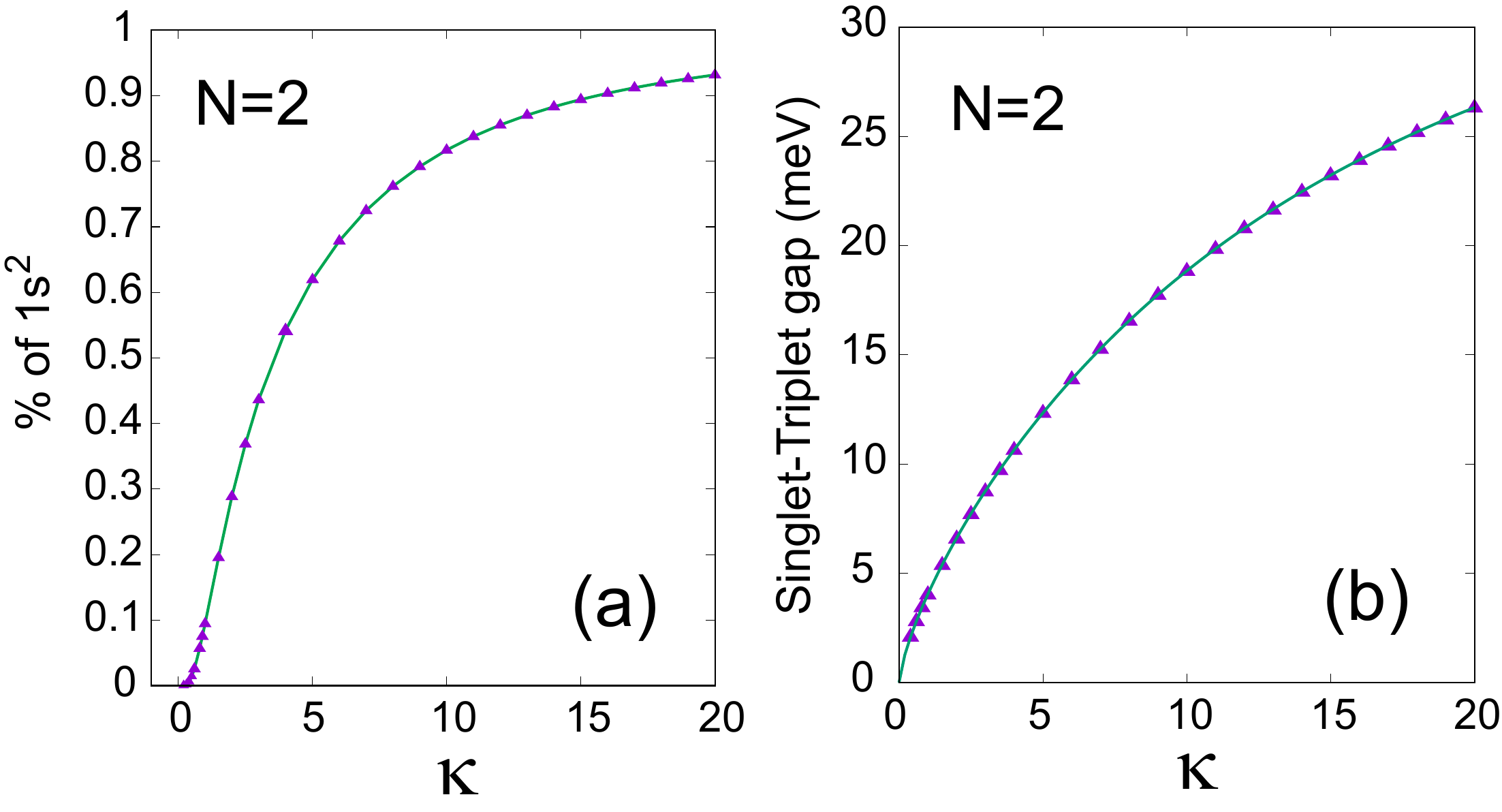}\caption{
(a) Weight (\%) of the $1s^2$ Slater determinant [an excellent approximation of the non-interacting ground 
state, see Fig.\ \ref{dens}(I)] in the FCI singlet ground state for $N=2$, as a function of $\kappa$.
(b) FCI singlet-triplet energy gap for $N=2$ as a function of $\kappa$. 
Remaining parameters: $a_M=14$ nm, $v_0=15$ meV, $\phi=10^\circ$, and 
$m^*=0.5m_e$  See the text for a full description.}
\label{aufb}
\end{figure}

Using the FCI calculations for $N=2$, we briefly discuss how the emergence of the Wigner molecularization 
negates the Aufbau principle that governs the natural atoms. Indeed, the ground-state for the non-interacting
MQD-Helium is approximated by an Aufbau-type single Slater determinant denoted as $1s^2$; see Fig.\ 
\ref{dens}(I). As shown in Fig.\ \ref{aufb}(a), the weight (\%) of the $1s^2$ Slater determinant in the FCI 
wave function decreases drastically with decreasing $\kappa$ (increasing correlations), and it 
vanishes for $\kappa \rightarrow 0$. In the CD plots, this behavior is associated with the 
developing of a depression at the origin (see Fig.\ \ref{dens}(a) and Fig.\ \ref{fig6} in the 
Appendix). 
For $\kappa \rightarrow 0$, singlet and triplet CDs coincide. Furthermore, in 
accordance, Fig.\ \ref{aufb}(b) demonstrates that the singlet-triplet energy gap [i.e., the gap 
$\Delta_{\rm GLE}$ between the ground state and the lowest excited one] decreases as well with decreasing 
$\kappa$, and it vanishes for $\kappa \rightarrow 0$. On the contrary, because $N=2$ is a 2D shell closure, 
the Aufbau approach would have yielded a value very close to the harmonic-oscillator gap, 
$\hbar \omega_0$ (i.e., 42.59 meV for the parameters used in Fig.\ \ref{aufb}).

The voiding of the Aufbau principle with decreasing $\kappa$ (increasing $R_W$) can be also seen from the 
quenching of $\Delta_{\rm GLE}$ for $N=6$, which is also a closed shell in the non-interacting limit.
Indeed, for $\kappa=3$ and same other parameters as in 
Fig.\ \ref{aufb}, we find values of 1.68 meV, 2.50 meV, and 4.61 meV for the $(S=0 \rightarrow S=2)$, 
$(S=0 \rightarrow S=1)$, and $(S=0 \rightarrow S=3)$ gaps, respectively. These values are to be contrasted 
with the non-interacting $\Delta_{\rm GLE}$ of $\approx \hbar \omega_0 = 42.59$ meV. This strong quenching of 
$\Delta_{\rm GLE}$'s conforms with the expected full degeneracy of Wigner crystalline states belonging to a 
given total-spin multiplicity in the ``classical'' limit $\kappa \rightarrow 0$
\cite{ [{A similar degeneracy happens also for contact interaction; see }] [{}]yann16}.  

The FCI second differences, $\Delta_2(N)=E(N+1)+E(N-1)-2E(N)$, associated with 
the ground-state energies, relate to the electrochemical potential gap $\Delta \mu$
\cite{kouw91}, which is experimentally accessible through measurements of the bilayer-sample capacitance 
\cite{mak21}. With respect to the TMD bilayers, the $\Delta_2(N)$ for the isolated MQD relates to capacitance
measurements at integer fillings $\nu \geq 1$. We found the FCI values: $\Delta_2(2)=93.55$ meV, 
$\Delta_2(3)=70.73$ meV, $\Delta_2(4)=78.06$ meV, and $\Delta_2(5)=60.34$ meV. These values suggest two FCI 
trends that are in general agreement with the results of Ref.\ \cite{mak21}: (i) Due to the Wigner 
molecularization, the $\Delta_2(N)$ values are substantially smaller than the corresponding Hubbard gap,
$U=e^2\sqrt{\pi}/(\kappa l_0)=269.85$ meV, and (ii) on the average, $\Delta_2(N)$ decreases with increasing 
$N$ \footnote{The $\nu$ assignments in Ref.\ \cite{mak21} correspond to $N-1$ here.}.

{\it Conclusions:\/} We show that the physics of quantum WMs 
\cite{yann99,yann00,yann02.2,mikh02,yann03,szaf03,yann04,yann06.3,yann07,
yang07,umri07,yang08,yann22,yann22.2}, 
underlies at integer fillings $\nu > 1$ that of TMD moir\'{e} 
materials, which as artificial 2D materials are fast developing into a promising experimental platform 
\cite{mak21,feld22}, spawning the emergence of the new field of twistronics 
\cite{manz17,kaxi20,macd18,fu20,ange21}.
Our analysis, using both FCI CDs and CPDs for the single M QD, demonstrates that
the anisotropy of the moir\'{e} confinement imposes an immediately visible $C_3$ symmetry on the CDs, 
whilst the intrinsic polygonal-ring geometry of the WMs remains hidden and is revealed only using the CPDs. 
Notable exceptions are the cases of $N=3$ (at weak) and $N=4$ (at strong anisotropies) charge carriers where, 
due to intrinsic and extrinsic symmetry coincidence, a pinned (0,3) and a pinned (1,3) WM appears in the CDs, 
respectively; we note here that the discrete structural features of the WMs, having a nested polygonal 
sliding-ring motif, can be brought to light and accentuated via pinning induced by symmetry-perturbing 
influences, such as QD shape-distortions (via twist, strain, or gating) or other symmetry-breaking effects 
(e.g., impurities) \cite{yann99,yann06,yann00.2,yann11,yann20,yann21}.
Formation of the correlated MWMs results in strongly quenched energy gaps compared to those 
expected from both the Hubbard model and the Aufbau principle. Our FCI results provide essential benchmarks 
for developing future many-body computational methodologies, and in particular those based on machine
learning and artificial intelligence \cite{bench23}. 
The quantum WM phases predicted here can be experimentally verified using scanning tunneling microscopy for 
the CDs \cite{feen18,wang21} and scanning probe microscopy for the CPDs \cite{szaf13,west00}.\\      

{\bf NOTE ADDED.}
We recently became aware of a preprint by D. Luo {\it et al.} 
\cite{[{See version 2 of~}][{}]fu23} (titled “Artificial intelligence for artificial materials: moiré atom”),
where Wigner molecularization in isolated TMD moir\'{e} QDs has been
addressed. The results reported in that eprint differ substantially from our benchmark FCI results. We note
in particular the inability of that study, based on neural network methodology, to (1) identify formation of
a Wigner molecule for $N=2$, and (2) recover all the exact charge densities for $N>3$. In the context of
point (2) above, we call attention to pertinent remarks in Ref.\ \cite{foul23} about ongoing issues
in fermionic neural networks pertaining to Wigner crystallization and the conservation of the symmetries of
the many-body Hamiltonian, which is a sine qua non for exact many-body solutions.\\

{\it Acknowledgments:\/}
This work has been supported by a grant from the Air Force Office of Scientific Research (AFOSR) under Award
No. FA9550-21-1-0198. Calculations were carried out at the GATECH Center for Computational Materials Science.

\appendix

\begin{figure*}[t]
\centering\includegraphics[width=16.0cm]{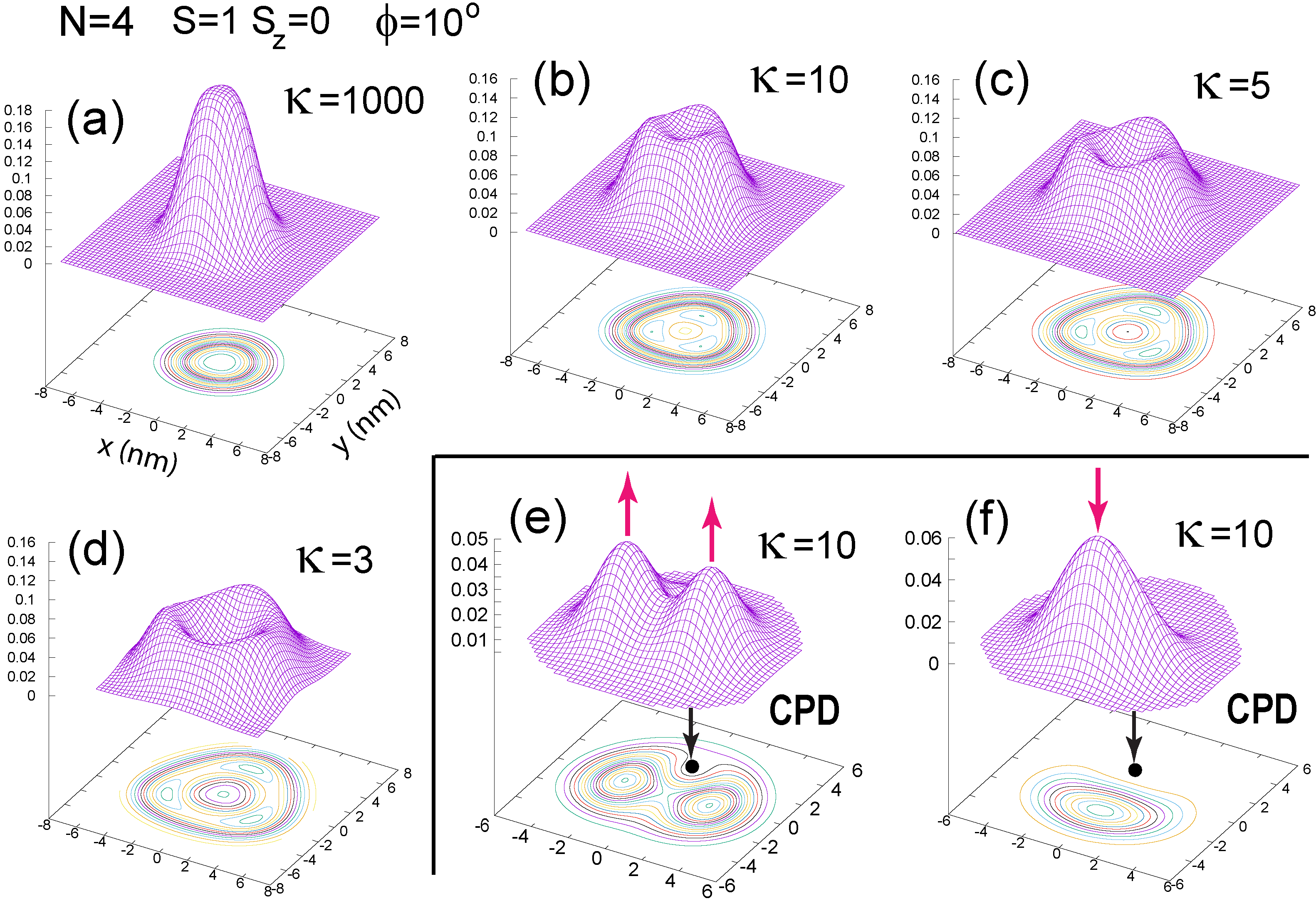}
\caption{
FCI moir\'{e}-QD charge densities for the total potential, $V_{\rm MQD}$, [Eq.\ (2) of the main text] for 
$N=4$ for various values of the dielectric constant $\kappa=1000.00$ (a), 10.0 (b), 5.0 (c), and 3.0 (d).
The (0,4) WM charge densities in frames (b), (c), and (d) contrast sharply with that of the 
non-interacting limit in frame (a). For $\kappa=10$, corresponding to a weakened $e-e$ repulsion compared 
to that induced by the hBN environment ($\kappa=5$), frames (e,f) provide further evidence of the 
unavoidability of WM formation in available TMD moir\'e materials. Remaining parameters: 
$S=1$, $S_z=0$, $m^*=0.5 m_e$, $a_M=14$ nm, $v_0=15$ meV, and $\phi=10^\circ$.}
\label{fig5}
\end{figure*}

\begin{figure*}[t]
\centering\includegraphics[width=16.0cm]{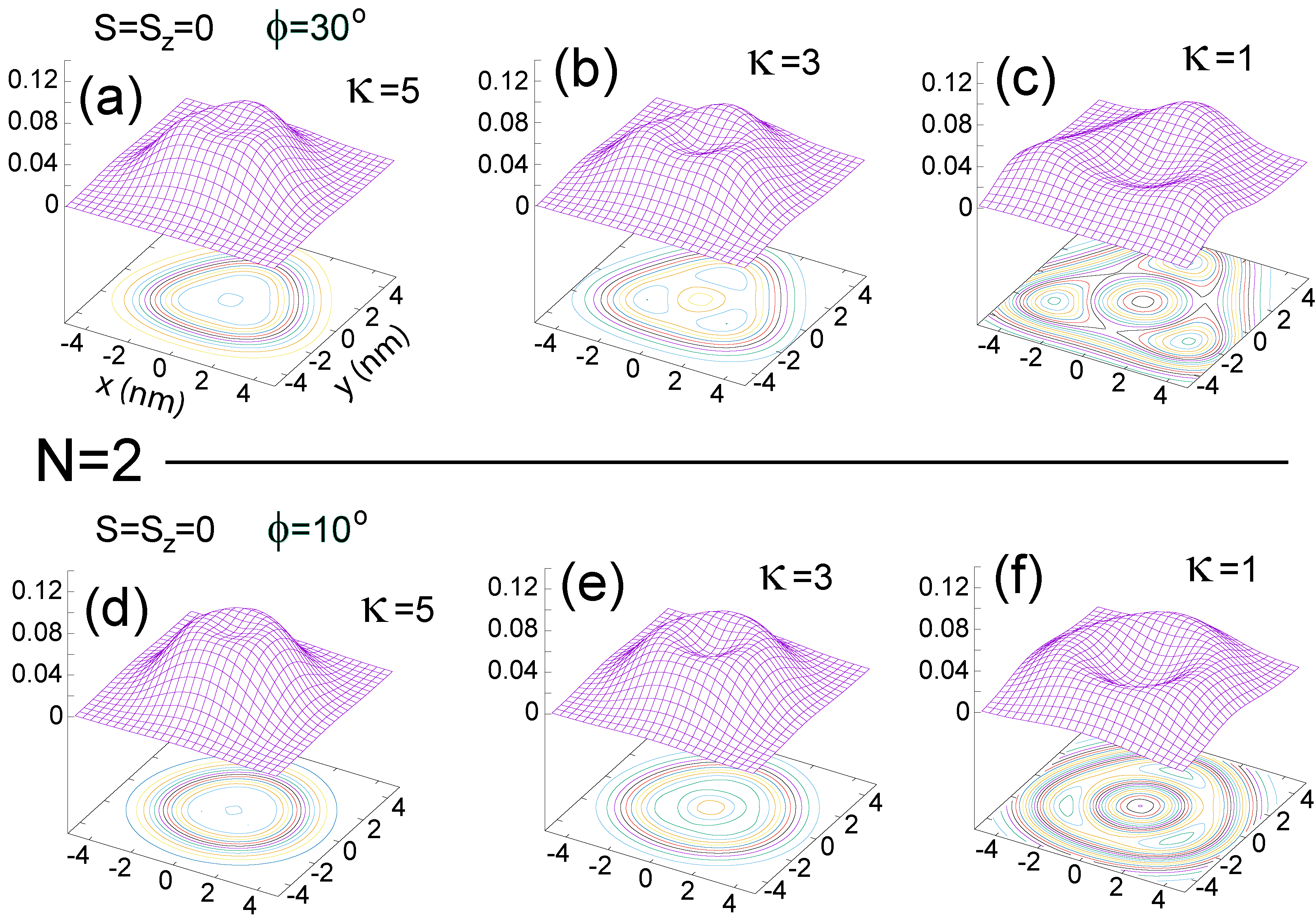}
\caption{
FCI moir\'{e}-QD charge densities [i.e., for the total potential, $V_{\rm MQD}$, in Eq.\ (2) of the main 
text which includes both the harmonic and cubic contributions] for the singlet state of $N=2$ holes, and 
for different values of the dielectric constant $\kappa$ and the angle $\phi$. Remaining 
parameters used: $a_M=14$ nm, $v_0=15$ meV, $m^*=0.5m_e$. (a-c) $\phi=30^\circ$. (d-f) $\phi=10^\circ$.
Left column: $\kappa=5$. Middle column: $\kappa=3$. Right column: $\kappa=1$. The corresponding Wigner 
parameter ($R_W$) values are: (a) 3.69, (b) 6.15, (c) 18.46, (d) 3.57, (e) 5.95, and (f) 17.85. The 
progressive enhancement of the depression at the origin as a function of decreasing $\kappa$
(increasing $R_W$) is clearly seen for the cases of both angles $\phi$. 
These ring-like charge densities correspond to a rotating Wigner molecule whose intrinsic geometric
configuration [denoted by (0,2)] consists of two antipodal fermions; see main text. CDs in units of
$1/nm^2$.}
\label{fig6}
\end{figure*}

\begin{figure*}[t]
\centering\includegraphics[width=16.0cm]{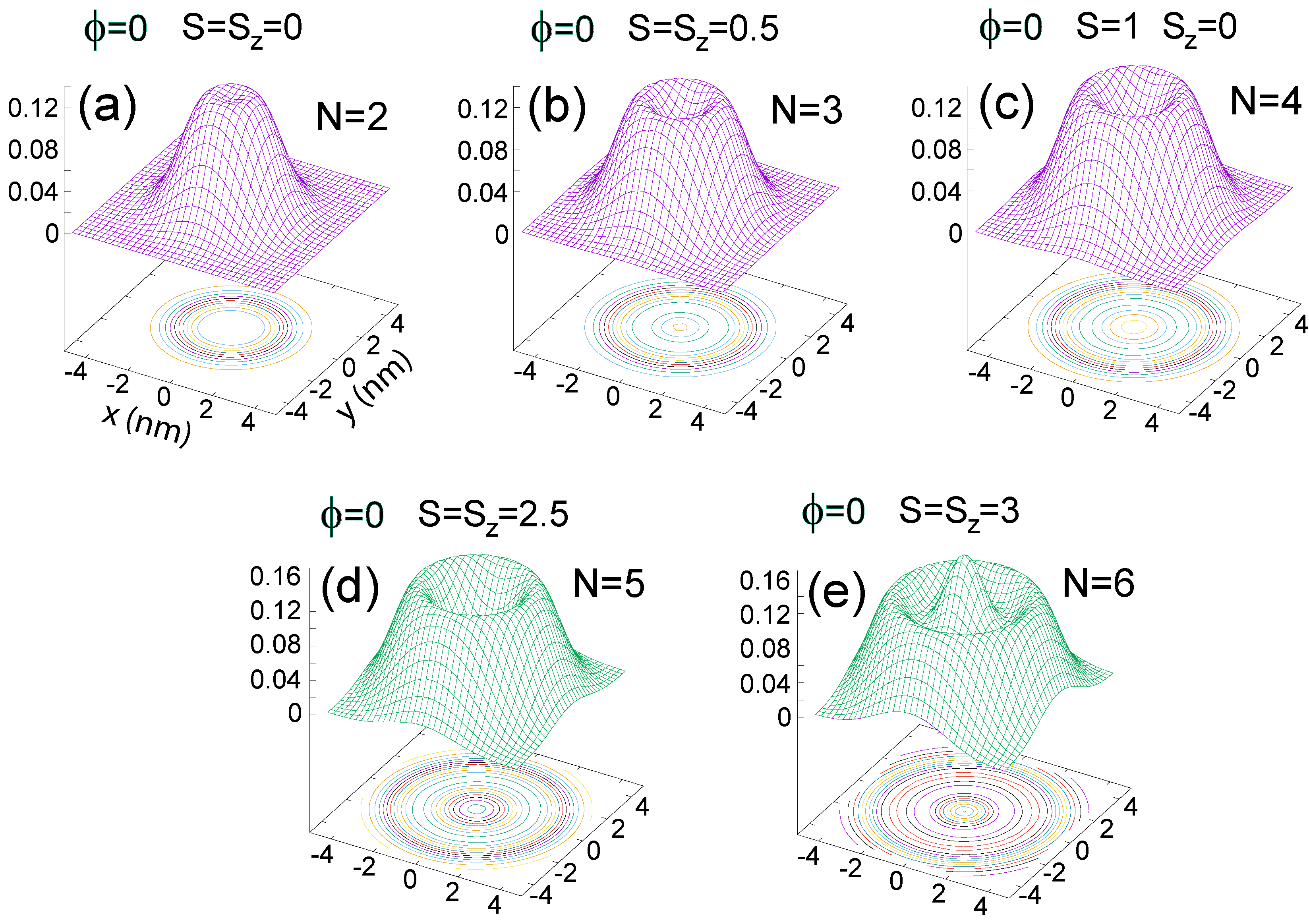}
\caption{
FCI moir\'{e}-QD, charge densities for the total potential, $V_{\rm MQD}$, [Eq.\ (2) of the main text] for a 
vanishing angle $\phi=0^\circ$. In this case only the harmonic term in the expansion survives, 
and the confinement has circular symmetry. (a) ground-state CD for $N=2$ holes. (b) ground-state CD for
$N=3$ holes. (c) ground-state CD for $N=4$ holes. (d,e) CDs of the lowest-in-energy fully spin-polarized 
states for $N=5$ and $N=6$ holes, respectively. 
The remaining parameters are: $a_M=15$ nm, $v_0=20$ meV, $m^*=1\;m_e$, and 
$\kappa=10$, which yields $R_W=2.88$. These uniform, ring-like charge densities correspond to a 
rotating Wigner molecule whose intrinsic geometric configuration [denoted as $(0,N)$ or $(1,N-1)$] consists 
of regular polygons with the particles being localized at the apices of the polygons. The results in this 
figure are in full agreement with the previous abundant literature on 2D semiconductor quantum dots; see
Ref.\ \cite{yann07} and references therein. CDs in units of $1/nm^2$.}
\label{fig7}
\end{figure*}

\section{THE CONFIGURATION INTERACTION METHOD}
\label{a1}

The full configuration interaction methodology has a long history, starting in quantum chemistry; see 
Refs.\ \cite{shav98,szabo}. The method was adapted to two dimensional problems and found extensive 
applications in the fields of semiconductor quantum dots
\cite{yann03,szaf03,ront06,yann07.2,yann07,yann22.2,yann22.3} and of the fractional quantum Hall effect 
\cite{yann04,jainbook}.

Our 2D FCI is described in our earlier publications. The reader will find a comprehensive exposition in
Appendix B of Ref.\ \cite{yann22.2}, where the method was applied to GaAs double-quantum-dot quantum 
computer qubits. The 
only difference with the present application to moir\'{e} QDs, due to the different external confinement, 
concerns the space orbitals, $\varphi_j(x,y)$, $j=1,2,\ldots,K$, that are employed in the building of the 
spin-dependent, single-particle basis used to construct the Slater determinants $\Psi^N_I$, which span 
the many-body Hilbert space [see Eq.\ (B4) in Ref.\ \cite{yann22.2}; the index $I$ counts the Slater 
determinants]. Indeed, for a moi\'{e} QD, the 
orbitals $\varphi_j(x,y)$ are determined as solutions (in Cartesian coordinates) of the auxiliary 
Hamiltonian
\begin{align}
H_{\rm aux} = \frac{{\bf p}^2}{2 m^*} +  \frac{1}{2} m^* \omega_0^2 (x^2+y^2),
\label{haux}
\end{align}   
where $m^*\omega_0^2=16 \pi^2 v_0 \cos(\phi)/a_M^2$, i.e., only the isotropic parabolic (harmonic) 
contribution in $V_{\rm MQD}$ is included.  

Following Ref.\ \cite{yann22.2}, we use a sparse-matrix eigensolver based on Implicitly Restarted 
Arnoldi methods (see ARPACK \cite{arpack}) to
diagonalize the many-body Hamiltonian in Eq.\ (3) of the main text. When a finite magnetic field
$B$ value is used, we replace ${\bf p}$ by ${\bf p}-(e/c){\bf A}(\br)$, where the vector potential
${\bf A}(\br) = 0.5B(-y,x)$ is taken according to the symmetric gauge.

The required one-body and two-body matrix elements for the ARPACK diagonalization are calculated as 
described in Ref.\ \cite{yann22.2}. In addition, the matrix elements 
$\langle \varphi_i(x,y) | \sin(3\theta) r^3 | \varphi_j(x,y) \rangle$
of the anisotropic term in the moir\'{e} confinement are calculated analytically using the algebraic
language MATHEMATICA \cite{math22} and the Hermite-to-Laguerre (Cartesian-to-polar) transformations listed
in Ref.\ \cite{kime93}.

In all calculations, we used at a minimum values of $K=45$ and/or $K=55$. For $N=2$ and $N=3$, additional 
calculations were carried out employing an extended single-particle basis with $K=78$. The maximum 
dimension of the many-body Hilbert space (the number of Slater determinants $\Psi^N_I$ in the FCI 
expansion) reached a value of $I_{\rm max} \approx 5,000,000$ in some cases.   

\section{CHARGE DENSITIES AND CONDITIONAL PROBABILITY DISTRIBUTIONS FROM FCI WAVE FUNCTIONS}
\label{a2}

The single-particle density (charge density) is the expectation value of a one-body operator
\begin{equation}
\rho({\bf r}) = \langle \Phi^{\rm FCI}_N
\vert  \sum_{i=1}^N \delta({\bf r}-{\bf r}_i)
\vert \Phi^{\rm FCI}_N \rangle,
\label{elden}
\end{equation}
where $\Phi^{\rm FCI}_N$ denotes the many-body (multi-determinantal) FCI wave function, namely,
\begin{equation}
\Phi^{\rm FCI}_{N} ({\bf r}_1, \ldots , {\bf r}_N) =
\sum_I C_I \Psi^N_I({\bf r}_1, \ldots , {\bf r}_N).
\label{mbwf}
\end{equation}

Naturally several distinct spin structures can correspond to the same charge density. The spin
structure associated with a specific FCI wave function can be determined with the help of the 
many-body spin-resolved CPDs \cite{yann00,yann04,yann06.3,yann07,yann09,yann15}.

The spin-resolved CPDs 
yield the conditional probability distribution of finding another fermion with up (or down) spin
$\sigma$ at a position ${\bf r}$, assuming that a given fermion with up (or down) spin $\sigma_0$ 
is fixed at ${\bf r_0}$. Specifically, the spin-resolved two-point anisotropic correlation function 
is defined as the the expectation value of a two-body operator
\begin{equation}
P_{\sigma\sigma_0}({\bf r}, {\bf r}_0)=
\langle \Phi^{\rm FCI}_N |
\sum_{i \neq j} \delta({\bf r} - {\bf r}_i) \delta({\bf r}_0 - {\bf r}_j)
\delta_{\sigma \sigma_i} \delta_{\sigma_0 \sigma_j}
|\Phi^{\rm FCI}_N \rangle.
\label{tpcorr}
\end{equation}

Using a normalization constant
\begin{equation}
{\cal N}(\sigma,\sigma_0,{\bf r}_0) =
\int P_{\sigma\sigma_0}({\bf r}, {\bf r}_0) d{\bf r},
\label{norm}
\end{equation}
we define a related conditional probability distribution (CPD) as
\begin{equation}
{\cal P}_{\sigma\sigma_0}({\bf r}, {\bf r}_0) =
P_{\sigma\sigma_0}({\bf r}, {\bf r}_0)/{\cal N}(\sigma,\sigma_0,{\bf r}_0),
\label{cpddef}
\end{equation}
having the property
$\int {\cal P}_{\sigma\sigma_0}({\bf r}, {\bf r}_0) d{\bf r} =1$.

\section{ADDITIONAL CHARGE DENSITIES AND CPDs FOR $N=4$ AND $\phi=10^\circ$}
\label{a3}

In Fig.\ \ref{fig5}, we compare the charge densities for $N=4$ at different values of the 
dielectric constant $\kappa$, i.e., $\kappa=1000.00$, 10.0, 5.0, and 3.0. We further display for $N=4$ the 
spin-resolved CPDs ${\cal P}_{\uparrow\downarrow}({\bf r}, {\bf r}_0)$ and 
${\cal P}_{\downarrow\downarrow}({\bf r}, {\bf r}_0)$. In all frames, the value of the anisotropy control 
parameter is set to $\phi=10^\circ$. For $\kappa=10$, corresponding to a weakened $e-e$ repulsion compared 
to that induced by the hBN environment ($\kappa=5$), frames (e,f) provide further evidence of the
unavoidability of WM formation in available TMD moir\'e materials.\\
~~~~\\

\section{ADDITIONAL CHARGE DENSITIES FOR $N=2-6$}
\label{a4}

In this part, we present additional charge densities for parameters, different from those
used in the main text; see Fig.\ \ref{fig6} and Fig.\ \ref{fig7}. In addition to the
angle value of $\phi=10^\circ$, charge densities for the values of $\phi=30^\circ$ and
$\phi=0^\circ$ are presented.\\


\nocite{*}
\bibliographystyle{apsrev4-2}
\bibliography{mycontrols,moire_WM_arx_v2}

\end{document}